# Quantification of nanoscale density fluctuations in hydrogenated amorphous silicon

Eike Gericke[1,2], Jimmy Melskens[3], Robert Wendt[1], Markus Wollgarten[4], Armin Hoell[1], and Klaus Lips[5,6]

[1]*Helmholtz-Zentrum Berlin, Institute for Nanospectroscopy, Hahn-Meitner-Platz 1, 14109 Berlin, Germany*
[2]*Humboldt-Universität zu Berlin, Department of Chemistry, Brook-Taylor-Str. 2, 12489 Berlin, Germany*
[3]*Eindhoven University of Technology, PO Box 513, 5600 MB Eindhoven, Netherlands*
[4]*Helmholtz-Zentrum Berlin, Institute Solar Fuels, Hahn-Meitner-Platz 1, 14109 Berlin, Germany*
[5]*Helmholtz-Zentrum Berlin, Department ASPIN, Hahn-Meitner-Platz 1, 14109 Berlin, Germany*
[6]*Freie Universität Berlin, Department of Physics, Arnimallee 14, 14195 Berlin, Germany*
e-mail address: lips@helmholtz-berlin.de

The nanostructure of hydrogenated amorphous silicon (a-Si:H) is studied by a combination of small-angle X-ray (SAXS) and neutron scattering (SANS) with a spatial resolution of 0.8 nm. The a-Si:H materials were deposited using a range of widely varied conditions and are representative for this class of materials. We identify two different phases which are embedded in the a-Si:H matrix and quantified both according to their scattering cross-sections. First, 1.2 nm sized voids (multivacancies with more than 10 missing atoms) which form a superlattice with 1.6 nm void-to-void distance are detected. The voids are found in concentrations as high as $6 \cdot 10^{19}$ cm$^{-3}$ in a-Si:H material that is deposited at a high rate. Second, dense ordered domains (DOD) that are depleted of hydrogen with 1 nm average diameter are found. The DOD tend to form 10-15 nm sized aggregates and are largely found in all a-Si:H materials considered here. These quantitative findings make it possible to understand the complex correlation between structure and electronic properties of a-Si:H and directly link them to the light-induced formation of defects. Finally, a structural model is derived, which verifies theoretical predictions about the nanostructure of a-Si:H.

The structure of amorphous materials such as hydrogenated amorphous silicon (a-Si:H) is often described through a continuous random network (CRN). In a CRN, the coordination of each network atom is equivalent to that of the corresponding crystalline lattice, but without having any long range order [1–3]. Such a structure is homogeneous on the nanoscale by definition and is expected to show only angle-independent elastic scattering in the range of small-angle scattering as it is described for pure liquids in literature [4]. However, experimental evidence of small-angle contributions in neutron and X-ray scattering as well as electron diffraction (SANS, SAXS, ED, respectively) for a-Si:H indicates that the CRN is not valid for the nanostructure of a-Si:H [3]. During the last two decades, considerable efforts have been made to resolve and understand the morphology of a-Si:H completely. This morphology is supposed to be related to the complex electronic, partially metastable properties of a-Si:H, such as the Staebler-Wronski effect (SWE) [5], or the glassy behavior of a-Si:H [1]. After over 50 years of intensive research on this topic we know today that the a-Si:H network is a complex mixture of amorphous and nano-sized crystalline-like domains in which nanovoids and vacancies of different sizes are embedded [3,6–13].

Until now, no experiments were conducted on a-Si:H materials to clearly resolve any feature smaller than 2 nm. Statements on smaller structures were only made on the basis of theoretical work. In such a work, Treacy and Borisenko recently reinterpreted experimental diffraction data. They assumed that a-Si:H consists of only two phases, namely a fully amorphous phase and topologically ordered domains [2]. Topologically ordered domains are referred to as nanostructures in the a-Si:H matrix that consists of six-membered rings of silicon only. A domain that provides such an environment must be at least 0.9 nm in size, have a high degree of order, and is hydrogen depleted. Note that such a topological structure can by definition still be amorphous.

The focus of our work is to lay the foundation for an improved understanding of the general a-Si:H nanostructure and its connection to the SWE. We will also verify the proposed theoretical model of Treacy and Borisenko, both by a combination of SAXS and SANS experiments [7,14–16]. For this purpose, we have prepared a series of a-Si:H samples with widely varying deposition conditions and distinctively different nanostructures. These samples represent an entire class of a-Si:H materials and were exceptionally well characterized by Fourier-transform infra-red (FTIR), positron annihilation spectroscopy (PAS), and electron spin resonance (ESR) [6–8,14,17]. We have employed SAXS and SANS, which provide a contrast on the electron density distribution (proportional to mass density) as well as the isotope distribution (primarily hydrogen density fluctuations), respectively. Due to improved detector sensitivity and the high brightness of the synchrotron radiation source used here, we achieved a real space resolution limit of 0.8 nm. This limit improves the limit of previously reported data by a factor of three [3]. From the scattering data, we derive a nanoscopic model of a-Si:H (fully amorphous, no crystalline inclusions) that consists of at least three domains: (i) a disordered a-Si:H matrix; (ii) dense ordered domains (DOD) which are fully amorphous, hydrogen depleted, and show a higher mass density than the a-Si:H matrix; and (iii) nm-sized voids of which the inner surface is decorated with hydrogen. The lower spatial resolution limit of our experiment was equal to the size of a vacancy of eight missing Si atoms in the a-Si:H matrix.

All a-Si:H samples were deposited on aluminum (Al) foil by radio-frequency plasma-enhanced chemical vapor deposition (rf-PECVD) from silane precursor gas. The a-Si:H nanostructure was altered by changing the deposition conditions by using well-known and established approaches, namely a low deposition rate (LDR, sample I and II), a high deposition rate (HDR, sample I and II) as well as a high hydrogen dilution (HHD, sample I and II) at a low deposition rate. HHD deposition was performed at a relatively high gas pressure of 8 mbar [18]. More details can be found in TABLE I in the supplementary information (SI) [19] and in Ref. [7,8,17]. The HHD and LDR material have a lower defect density when compared to the HDR material (see TABLE II in SI and Ref. [18]). To avoid significant X-ray attenuation by the Al foil, the a-Si:H film was separated from the Al substrate by an etching step (hydrochloric acid) and a subsequent cleaning in deionized water [19]. SAXS was measured at BESSY II at an X-ray energy of 9658 ± 2 eV (scattering vector (q) range: 0.06 to 7.7 nm$^{-1}$) [20–22]. SANS was measured at the V4 at the BER II reactor of Helmholtz-Zentrum Berlin (HZB) with wavelength of 0.45 and 0.60 nm ± 10.5 % (q-range: 0.1 - 6.7 nm$^{-1}$) [23,24]. For SANS, the a-Si:H films were measured directly on the Al foil, since the attenuation of the neutron beam by the Al is negligible. All SAXS and SANS data were normalized to absolute scattering cross-sections [25]. Transmission electron microscopy (TEM) and selected-area electron-diffraction (SAED) images were recorded using an energy-filtered Zeiss LIBRA 200FE microscope. TEM was employed in bright-field (BF-TEM) mode and hollow-cone illumination dark-field mode (HCDF-TEM).

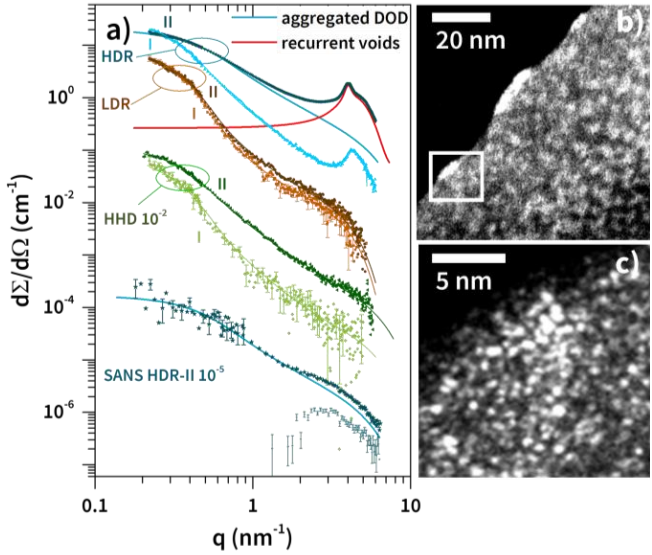

Fig. 1. (a) Experimental SAXS and SANS of the indicated samples and fits according to equations (4) (aggregated DOD) and (8) (recurrent voids) in the SI. The intensities have been shifted vertically by the indicated factors for better visibility. (b) HCDF-TEM and (c) fixed azimuthal HCDF-TEM of sample HDR-II taken in the indicated area in (b).

Fig. 1a shows SAXS and SANS data obtained on the LDR, HHD, and HDR a-Si:H sample series. The SAXS data are normalized and corrected for the contribution of surface roughness, angle-independent scattering, and the contribution of the broadened Si(111) reflection of the amorphous network for $q > 6\ nm^{-1}$, described in detail in the SI [19]. SAXS data were acquired up to a q-value of 7.7 nm$^{-1}$. However, for $q > 6.6\ nm^{-1}$ the data did not exceed the noise ratio. All SAXS results on the different samples are similar in shape, except for the HDR samples which exhibit a distinct and intense scattering for high q-values. The LDR and HHD series show two broad shoulders at $2 - 3\ nm^{-1}$ and at ~0.2 nm$^{-1}$. The shoulders indicate structures of about one and several nm in size, respectively. Nanostructures producing a shoulder at 0.2 nm$^{-1}$ were already discussed by Williamson et al. from SANS data [26]. In comparison, the HDR series yield an overall higher scattering intensity and exhibit the same broad shoulders as the LDR series. In addition, HDR samples show a distinct and intense scattering for high q-values with a pronounced peak at about 4 nm$^{-1}$, which has not been reported before and indicates a recurrent structure with a spacing of $d = 2\pi \cdot q^{-1} = 1.6\ nm$. The SANS data shown in Fig. 1a were evaluated in a similar fashion as SAXS. The shape of the SANS and SAXS curves of sample HDR-II differ significantly. The distinct peak at 4 nm$^{-1}$ is not visible in SANS in Fig. 1a. The data in Fig. 1a clearly show that we can resolve characteristic nanostructures in a-Si:H depending on the deposition conditions of the samples. The nanostructures are related to variations in the electron (SAXS) and isotope (SANS) density distribution in size ranges between 0.8 and 20 nm.

The BF-TEM (contrast: electron density) and HCDF-TEM (contrast: local atomic structure) images show LDR-I to be quite homogeneous (Fig. SI 3), where HDR-II shows distinctive structures, depicted in HCDF-TEM images in Fig. 1b,c. These images resolve a local electron diffraction cross-section contrast and thus the local ordering on a length scale of a few nm. In addition, in Fig. 1c HCDF-TEM is measured at a fixed azimuthal to discriminate the diffraction contributions for a certain alignment. The latter HCDF-TEM images clearly show domains of local ordering with below 1 nm size. Hence, we conclude that similar feature sizes are observed in TEM, SAXS, and SANS, probably related to the same nanoscopic structures.

To interpret the SAXS and SANS data of all samples, we consider four typical structural models which are (i) monodisperse spheres, (ii) polydisperse spheres, (iii) spheres forming aggregates, and (iv) spheres forming ordered superstructures. Simulated curves of these models are shown in Fig. 2 [19]. Each of these model structures influences the scattering curves in a very distinct and distinguishable manner. Since no crystalline inclusions are observable in TEM (Fig. SI 4), we assume that the Si films consist of two different amorphous phases with different atomic and electron densities leading to contrast in SAXS. Such a model was already proposed from experimental SANS and SAXS [26–28] and confirmed by theory [2] with domain sizes varying between 1 and 2 nm. If we assume that such dense domains are monodisperse spheres with radius $R$ embedded in an amorphous phase, they will produce a scattering curve displayed in Fig. 2a with interference fringes well known for the Fourier transformation to reciprocal space and a scattering cross-section ($d\Sigma/d\Omega$, integrated differential scattering cross-section per unit solid angle), which is a measure of the specific scattering contrast and the scattering volume. The blue arrow in Fig. 2a indicates the q-value which is associated with a sphere of diameter $2 \cdot R$ through the expression $1/q \approx R$. If, instead, the spheres have a size distribution (polydisperse), the fringes in Fig. 2a are smoothed out but the general shape of the scattering curve remains. If these spherical domains randomly agglomerate this will generate an additional shoulder at lower $q$-values as indicated by the green arrow in Fig. 2c, determining the average agglomerate radius $R_g^a$. If the polydisperse spheres cluster in smaller groups of two or more spheres with a distance $d$, as indicated in Fig. 2d, this superstructure will then dominate the complete scattering curve. The scattering intensity of the single spheres is reduced to the benefit of an arising peak at $q = 2\pi/d$ (indicated by the red arrow), as discussed elsewhere [29]. The detailed procedure on the calculation of these model scattering curves can be found in the SI [19] and are well established in the SAXS and SANS literature [30].

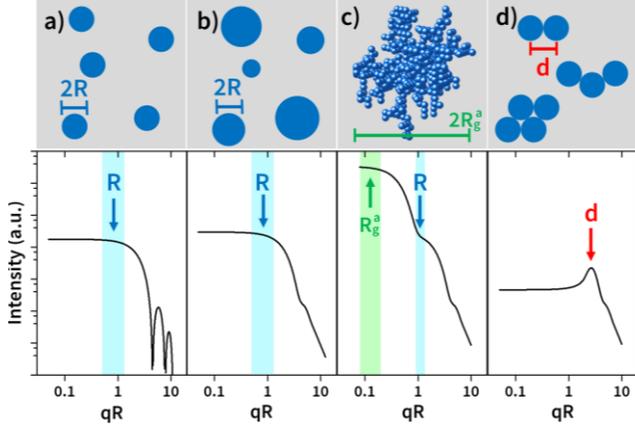

Fig. 2. Model structures and calculated scattering curves of monodisperse particles (a), polydisperse (20 %) particles (b), random aggregated particles (c), and particles clustered with a recurrent distance (d) embedded in an amorphous phase. For simplicity, a generalized axes $qR$ is shown.

To theoretically reproduce the scattering curves displayed in Fig. 1a, we have used standard scattering theory as described above to fit the experimental data. The TEM image in Fig. 1c clearly indicates that a DOD with a typical diameter of 1 nm exists in the amorphous matrix, forming agglomerates of a few nm in size (Fig. 1b). In the following, we assume the existence of the structures identified in TEM: (i) fully amorphous silicon; (ii) DOD; (iii) aggregated DOD. However, to be able to describe the scattering curves of HDR samples fully, we need an additional domain (iv), which forms a superstructure of a yet unidentified phase originating in the peak at 4 nm⁻¹ in Fig. 1a. For simplicity, we assume a spherical and polydisperse character for all phases. Small deviations from the spherical character only slightly alter the scattering curves and will not lead to noticeable changes outside the error margin of the scattering curves [30].

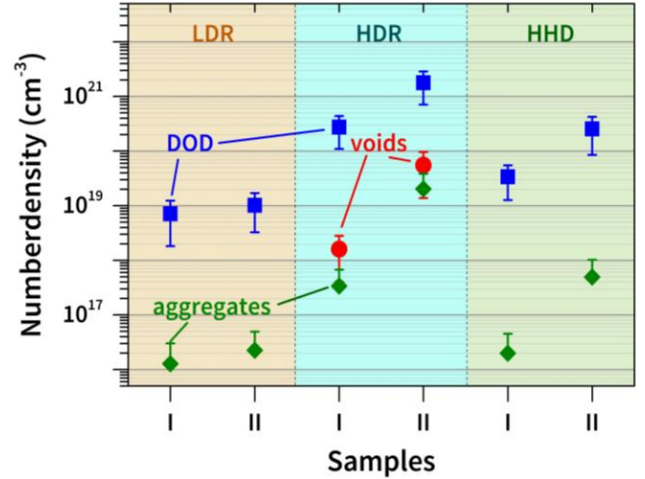

Fig. 3. Number density of DOD, voids, and aggregates in the studied samples. The number density of Si atoms in crystalline silicon is $5 \cdot 10^{22} cm^{-3}$.

Fits to the data using equations (4) and (8) from the SI [19] are shown in Fig. 1a (solid lines). The model of agglomerated DOD (4) reproduces the experimental SAXS curves for the LDR and HHD sample series very well. The model predicts DOD with a diameter of $1.3 \pm 0.2\ nm$ (LDR) and $0.8 \pm 0.1\ nm$ (HHD) samples forming agglomerates with diameters $2 R_g^a \approx 10 - 15\ nm$. The distinct peak at $q = 4\ nm^{-1}$ in the HDR series can be reproduced with the model of ordered superstructures. As discussed above, this peak position indicates a recurrent distance of $d = 1.6\ nm$. The peak shape indicates a diameter of about $1.2 \pm 0.3\ nm$ of the underlying spherical phases. The size of the individual cluster of the superstructure cannot be determined from our scattering data.

We speculate that the superstructure-forming phase is due to small voids of which the inner surfaces are decorated by hydrogen. To prove this, we provide counterevidence and fit the SANS data of the HDR-II sample using the agglomerated DOD model alone. Since the scattering contrast in SAXS and SANS depends on fundamentally different physical properties, SANS will have a clearly smaller contrast concerning the voids. However, SANS has a strong contrast for hydrogen and is hence very sensitive to distinguish a hydrogenated from a hydrogen-depleted silicon phase. Consequently, the contrast of DOD embedded in the amorphous phase for SAXS is $\Delta\eta^2 = 0.069 \cdot 10^{21} cm^{-4}$ (electron density contrast) and for SANS (H-distribution contrast) is $\Delta\eta^2 = 1.35 \cdot 10^{21} cm^{-4}$. Using these contrast values, we have fitted the experimental curves and find that the superstructure in SAXS nearly vanishes in the SANS data of sample HDR-II. Hence, this is proof that the superstructure cannot be associated with DOD but instead with nm-sized voids in the amorphous matrix. H-depleted and H-filled voids will produce two orders of magnitude higher contrast in SAXS ($161 \cdot 10^{21}\ cm^{-4}$) compared to SANS ($1.76 \cdot 10^{21}\ cm^{-4}$), respectively. Fits with H-filled voids to the SANS data are displayed in Fig. 1a and show good agreement with the experimental data. The fact that the residual of the SANS fit shows a component at $2 – 6$ nm$^{-1}$ can be taken as an indication that the voids are H-filled. Note that we were not able to reliably fit the experimental SANS data with such a H-filled void model due to large uncertainties in the local model parameters.

Since all SAXS and SANS experiments were measured with calibration of the absolute scattering cross-sections, we obtain quantitative information about the number density of the various phases identified in the a-Si:H samples, as shown in Fig. 3. The number density of individual voids within the superstructure is between $10^{19}$ and $10^{20}$ cm$^{-3}$ and these voids are only found in HDR samples. The number density of the DOD strongly depends on the deposition conditions and range from $10^{19}$ cm$^{-3}$ in LDR to $10^{21}$ cm$^{-3}$ in HDR samples. The DOD always form aggregates, which can be found in a number density proportional to the DOD, Fig. 3. We arrive at the conclusion that LDR material always has a low concentration of DOD and no nanovoids, which indicates a homogeneous material. This finding correlates to observations from FTIR spectroscopy, indicating LDR materials to be dominated by a strong 2000 cm$^{-1}$ absorption mode [17,18]. In contrast, a high nanovoid density is always associated with a very high density of DOD and is found in HDR material. Again, the correlation with FTIR findings is impressive, which describes HDR material as porous and dominated by a strong 2100 cm$^{-1}$ absorption mode [17,18]. If LDR materials contain voids, either the size of the voids was below the resolution limit, or their number density was too low to exceed the statistical noise.

Our experimental findings lead us to conclude that a-Si:H has a nanostructure as depicted in Fig. 4 that consists of at least three clearly distinguishable phases depending on the deposition conditions and associated material properties. (i) Voids of 1.2 nm diameter clustered with a recurrent distance of 1.6 nm are only found in porous a-Si:H deposited at a high deposition rate. (ii) DOD are found in all samples considered in this study and have diameters of $0.8 \pm 0.1\ nm$ for HDR and HHD series and $1.3 \pm 0.2\ nm$ for LDR series. DOD always form aggregates of a few hundred to thousand DODs per aggregate and a diameter of $2\ R_g^a \approx 10 - 15\ nm$. Both phases, voids and DODs are embedded in (iii) the a-Si:H matrix phase. Note that the resolution of our experiment makes it possible to resolve voids of a size equivalent to 8 missing atoms in the Si network. We find that voids appear as single nanovoids in 50-60 % of the cases but frequently cluster in a di- or multi-void superlattice structure (red phase in Fig. 4) while the inner surfaces of the nanovoids are decorated with hydrogen. We believe that such superstructures form through the diffusion of nanovoids during growth and release the stress in the amorphous network. The density of nanovoids is correlated with the density of DOD. The correlation is not resolved for samples with low DOD number density such as the LDR, possibly due to signal-to-noise and q-range limitations. We also find that the DOD density is correlated with the density of paramagnetic defects in the a-Si:H material. Much more impressive is the fact that clustered voids are found with the same spacing of about 1.6 nm as clustered paramagnetic defects [14]. These clustered defects are generated during light-induced degradation and are part of the SWE [8]. It can be speculated that these defects are generated and stabilized in neighboring voids. Hence, we strongly suggest that the structural model presented here is directly linked to the light-induced formation of defects and this view on the nanostructure has to be considered to describe the complex dynamics of the SWE. Unfortunately, with the resolution of our experiments we are currently not able to resolve divacancies as reported by Smets and et al. [15], which are suggested to be directly linked to the degradation of electronic properties of a-Si:H.

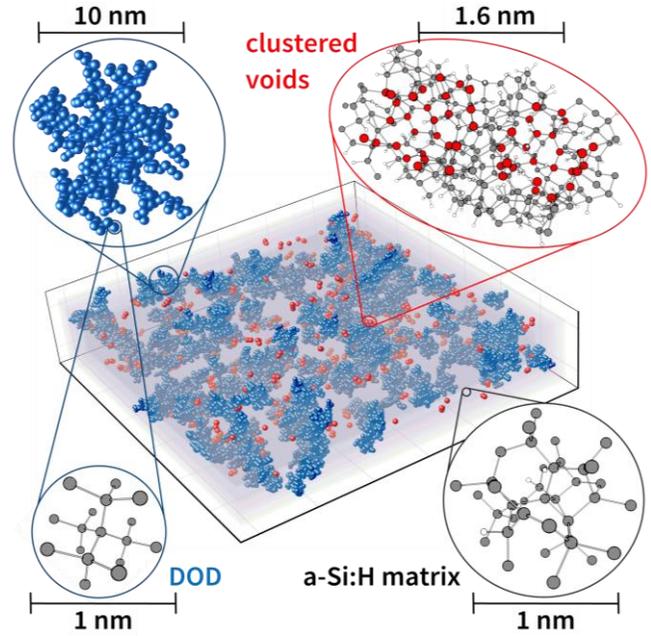

Fig. 4. Nanoscopic structure model of HDR a-Si:H according to the scattering and TEM data. DOD and voids are shown as blue and red spheres, respectively. The a-Si:H matrix is depicted as a haze of gray. Closeups show specific atomic models (Si atoms grey, H atoms white, Si atoms at the void surface in red). Atomic models were calculated by DFT and semiempirical methods. Coordinates for the nanoscopic model are generated by a python script available under GNU GPL v3.0 license at
https://github.com/EikeGericke/3D-Model-of-amorphous-silicon

In summary, we identify voids of about 1.2 nm in size which appear to be clustered to multi-void superstructures with 1.6 nm recurrent distance. These voids are expected to play a crucial role in the SWE. In addition, our experimental findings successfully verify the theoretical predictions of a-Si:H being a two-phase material (one phase being hydrogen-depleted) [2] for a broad range of different a-Si:H sample morphologies and show that the material can clearly not be represented by a CRN.

## ACKNOWLEDGEMENTS

All SAXS measurements were performed in cooperation with the Physikalisch-Technische Bundesanstalt (PTB) at the PTB-FCM beamline at BESSY II where we thank Michael Krumrey and the operational beamline staff. We gratefully acknowledge Uwe Keiderling for experimental support at the SANS instrument V4 as well as in the preparation of the manuscript, U. Bloeck (Corelab Correlative Microscopy and Spectroscopy CCMS, HZB) for the preparation of the TEM sample, and Dragomir Tatchev for assistance in data evaluation. The laboratory infrastructure of Delft University of Technology is acknowledged for the fabrication of the a-Si:H samples.